\documentclass[prl,aps,twocolumn,showpacs,floatfix,groupedaddress]{revtex4-1}

\usepackage{graphicx}% Include figure files
\usepackage{dcolumn}% Align table columns on decimal point
\usepackage{bm}% bold math
\usepackage{latexsym}
\usepackage{amsmath}
\usepackage{amssymb}
\usepackage[latin1]{inputenc}
\usepackage[usenames]{color}
\usepackage{wasysym}
\usepackage[usenames,dvipsnames]{xcolor}
\usepackage{multirow}

\begin{document}

\title{Surface depression with double-angle geometry during the discharge \\ of close-packed grains from a silo.}

\author{F. Pacheco-V\'azquez$^*$, A.Y. Ramos-Reyes and S. Hidalgo-Caballero}

\affiliation{Instituto de F\'isica, Benem\'erita Universidad Aut\'onoma de Puebla, Apartado Postal J-48, Puebla 72570, Mexico}
\date{\today}
%\date{February 25, 2009}

\begin{abstract}

When rough grains in standard packing conditions are discharged from a silo, a conical depression with a single slope is formed at the surface. We observed that the increase of the volume fraction generates  a more complex depression characterized by two angles of discharge: a lower angle close to the one measured for standard packing and a considerably larger upper angle. The change in slope appears at the boundary between a densely packed stagnant region at the periphery and the central flowing channel formed over the aperture. Since the material in the latter zone is always fluidized, the flow rate is unaffected by the initial packing of the bed. On the other hand, the contrast between both angles is markedly smaller when smooth particles of the same size and density are used, which reveals that high volume fraction and friction must combine to produce the observed geometry. Our results show that the surface profile helps to identify by simple visual inspection the packing conditions of a granular bed, and this can be useful to prevent undesirable collapses during silo discharge in industry.

\end{abstract}

%\pacs{45.70.-n, 45.50.-j}

\maketitle

\section{I. INTRODUCTION}

The discharge of dry grains from a container reveals several features of a granular material depending on the observation zone: if we look at the bottom, the discharged material accumulates forming a conical heap, the grains roll and slide in continuous or intermittent avalanches until the lateral face reaches the angle of repose $\theta_R$\cite{Duran2000}; if we look at the aperture, we observe a constant flow of grains $Q$ independent of the height of the granular column\cite{Beverloo1961}, but we can also observe jamming if the opening approaches to the grain size\cite{Zuriguel2005,Zuriguel2014}. Inside the container, on the other hand, the motion of material towards the aperture can develop three different patterns: mass-flow, funnel-flow or intermediate flow\cite{Tejchman2013,Nguyen1980,Drescher1992,Grudzien2015}. In the first case, typically observed in hoppers with very inclined walls, the material moves toward the outlet in a uniform manner producing a continuous descent of the surface level; in the funnel flow, which is observed in silos with an opening through the horizontal bottom wall, an active flow channel forms above the outlet with stagnant material at the periphery; the third case of flow occurs when there is a combination of funnel flow in the lower part and mass flow in the higher part of the granular column. Flow pattern is fundamental to any understanding of stresses acting on the granular material, and modern techniques have been applied to visualize it, including particle image velocimetry, electrical capacitance tomography and X-ray tomography\cite{Babout2013,Tejchman2013,Grudzien2015}). Now, if we finally observe at the upper surface of the column, a conical depression may appear, with slope slightly larger than the one corresponding to the maximum angle of stability $\theta_M$, namely, the angle at which the material starts to flow\cite{Albert1997,Boumans2012}. 

Flow rate, flow patterns, $\theta_M$ and $\theta_R$ depend on different factors like particle shape and size \cite{Robinson2002, Olson2002, Arias2011}, rolling and sliding friction coefficients \cite{Zhou2002}, roughness \cite{Pohlman2006} wall separation\cite{DuPont2003}, container and aperture size\cite{Beverloo1961,Mankoc2007}, polydispersity\cite{Pournin2007}, volume fraction\cite{Olson2005,Unac2012,Babout2013,Aguirre2014}, and other complex parameters like liquid content \cite{Nowak2005,Scheel2008,Pacheco2012} and acceleration of gravity \cite{Kleinhans2011,Dorbolo2013}. It is well known that $Q$ is mainly determined by the outlet-diameter-to-grain-size ratio and kind of grains. Since the pioneering expression reported by Beverloo in that regard\cite{Beverloo1961}, more accurate models were lately proposed to include intermittent flow produced by jamming \cite{Mankoc2007,Zuriguel2014}. More recently, the discharge of granular materials have been investigated using submerged silos\cite{Wilson2014,Koivisto2016} or particles with repulsive magnetic interactions\cite{Lumay2015,Hernandez2017}.

%%%%%%%%%%%%%%%%%%%%%%%%%%%%%%%%%%%%%%%%%%%%%%%%%%%%%%%%%%%%%%%%%%%%%%%%%%%%%%%
\begin{figure*}[ht!]
\begin{center}
\includegraphics[width=14cm]{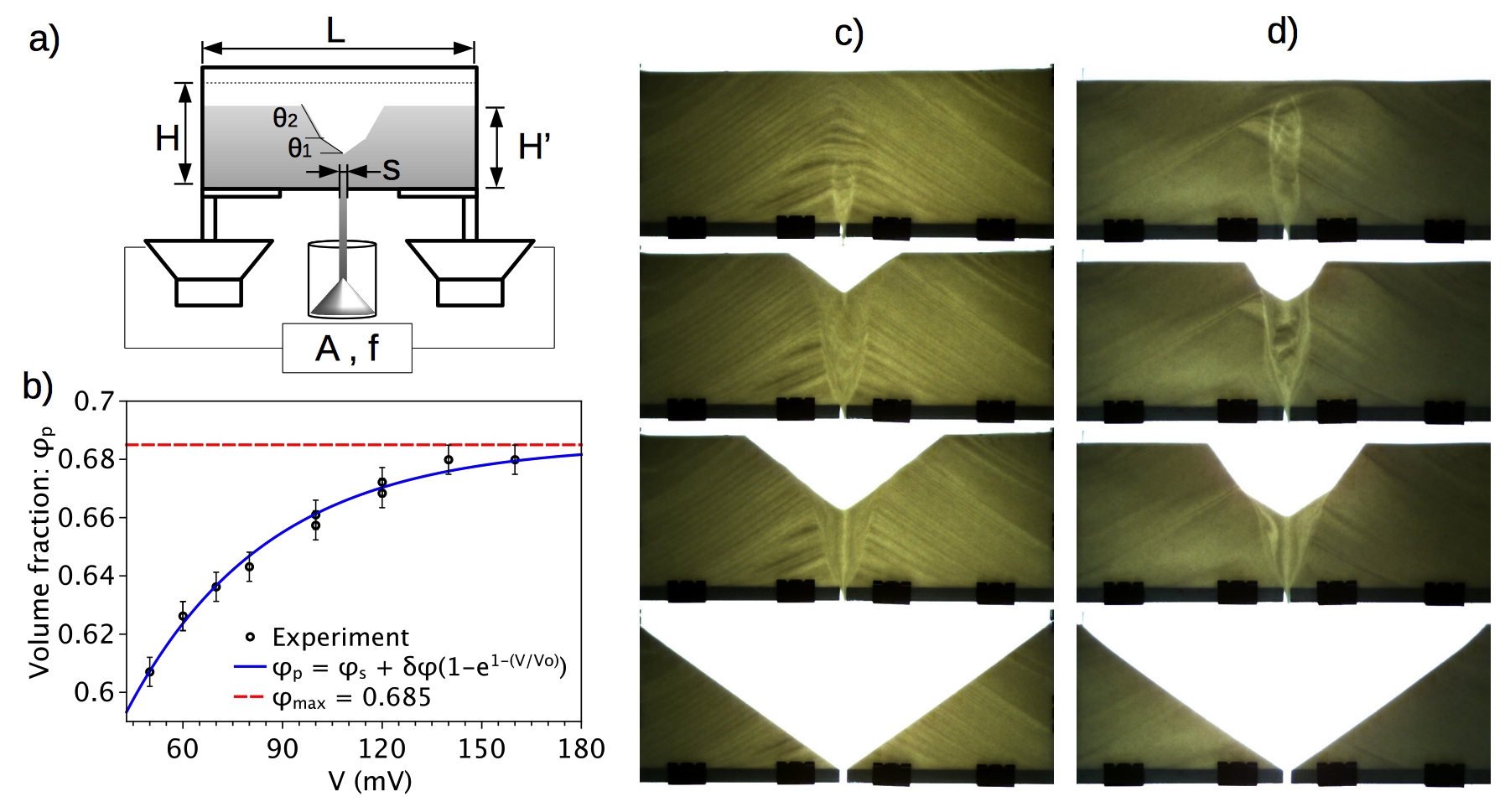}
\caption{Two-dimensional experiment: a) Experimental setup. b) Calibration curve indicating the packing fraction $\varphi_p$ obtained when the cell is vibrated at different vibration amplitudes $V$. c-d) Snapshots of the discharge of sand with c) standard packing $\varphi_s=0.59$, and d) higher packing $\varphi_p =0.66$. Two external angles are observed in the second case.}
\label{fig1}
\end{center}
\end{figure*}
% %%%%%%%%%%%%%%%%%%%%%%%%%%%%%%%%%%%%%%%%%%%%%%%%%%%%%%%%%%%%%%%%%%%%%%%%%%%%%%%  

Although several investigations concerning silo discharge have been focused on the flow rate, the pile formation and in the features of the flow pattern, the description of the upper depression and its evolution during the discharge has gone unnoticed. Here we report that, if a silo filled with sand is vibrated or tapped before the discharge, the flow rate is not affected by the increase of packing fraction but the surface depression suffers an important change in geometry. The most remarkable effect is the observation of two external angles that contrast with the constant slope of conical depressions typically reported in the literature for standard packing conditions\cite{Boumans2012,Wilson2014,Loranca2015}. We found that these two angles appear because of the existence of two regions with distinct packing fractions: a peripheral stagnant zone and a less dense central zone flowing towards the aperture. Experiments in two-dimensional (2D) and three-dimensional (3D) silos were developed and the dynamic evolution of the depression was measured. On the other hand, a less pronounced change in slope was observed when spherical and smooth glass beads of the same size were used. This indicates that friction increases markedly due to the greater contact area among amorphous particles at high volume fractions, and such combination produces the double angle geometry.

\section{II. Experimental setup}

The 2D system consists of a Hele-Shaw cell of inner dimensions $l=28.00 \times H= 10.05 \times w= 0.31$ cm$^3$ built with transparent glass walls of 0.5 cm thick separated by aluminium bars, and with a rectangular opening of $0.36 \times 0.31$ cm$^2$ centrally located at the bottom. The container was filled with a mass $m = 136.4 \pm 0.1$ g of sand grains (density $\rho = 2.65\pm0.01$ g/cc and size distribution of $150-250$ $\mu$m) poured gently from the top to obtain a standard volume (packing) fraction $\varphi_s = m/l H w \rho \approx 0.59 \pm 0.01$. The cell was vertically located on a horizontal aluminium base mounted on two subwoofers controlled with a function generator which allowed us to apply a sinusoidal vibration of different voltage amplitudes $V$ at a fixed frequency $f=60$ Hz, see fig. \ref{fig1}a. With this frequency, the sand level decreased uniformly along the cell and the vibration was maintained until the sand surface reached a constant level $H'$. The larger the applied voltage, the larger was the shift in the sand level, revealing a higher volume fraction $\varphi_p = \varphi_s H/H'$ ranged in the interval $0.59 < \varphi_p < 0.68$. The dependence $\varphi_p$ vs $V$ was well fitted by the equation $\varphi_p= \varphi_s + \delta \varphi (1-e^{-V/V_o})$, where $\varphi_s =0.59 \pm 0.01$ is the standard packing obtained without vibration, $\delta \varphi =\varphi_{max} -\varphi_s =0.093$, and $V_0=42.4$ mV  is the minimum voltage necessary to increase the packing of the bed, see fig. \ref{fig1}b. From the data fitting, it was also found that the maximum packing fraction reached by this granular material is $\varphi_{max} = 0.685 \pm 0.005$. After the compaction process, the flow was started and the discharge was filmed at 125 fps with a high speed camera Photron SA3. The process was repeated five times for a certain value of $\phi$ and the videos were analysed using ImageJ to measure the surface angles and total time of discharge. An important advantage of the 2D version is that we were able to observe the flow pattern during the whole discharge.

%%%%%%%%%%%%%%%%%%%%%%%%%%%%%%%%%%%%%%%%%%%%%%%%%%%%%%%%%%%%%%%%%%%%%%%%%%%%%%%

\begin{figure*}[ht!]
\begin{center}
\includegraphics[width=17cm]{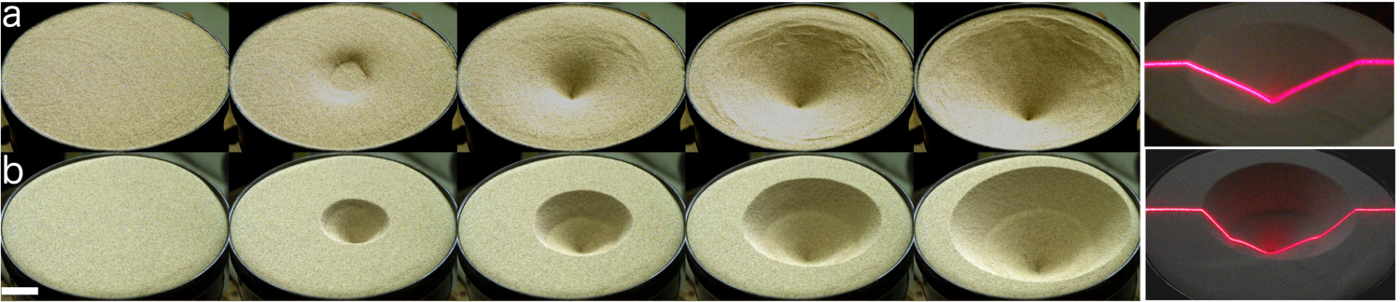}
\caption{Three-dimensional experiment: a) Conical depression observed during the discharge of sand with packing fraction $\varphi_s$. b) Depression with two surface angles observed when the same material with $\varphi_p > \varphi_s$ is discharged. In the last two snapshots a laser line was used to better visualize the surface profile.
}
\label{fig2}
\end{center}
\end{figure*}
% %%%%%%%%%%%%%%%%%%%%%%%%%%%%%%%%%%%%%%%%%%%%%%%%%%%%%%%%%%%%%%%%%%%%%%%%%%%%%%%

In the 3D case, a brass cylindrical container of 29 cm inner diameter and 30 cm height with a circular aperture of $1.5\pm0.1$ cm radius at the bottom face was filled with approximately 31.5 kg of sand. Due to the weight of this system, it was impossible to vibrate; then it was tapped laterally with a rubber hammer to increase the packing fraction. This method was less reproducible but allowed us to observe the equivalent double angle geometry found in the 2D case. To measure the depression profile, a laser line was projected centrally on the sand surface during the discharge and it was recorded at 60 Hz with a lateral view at 45$^{\circ}$.

 %%%%%%%%%%%%%%%%%%%%%%%%%%%%%%%%%%%%%%%%%%%%%%%%%%%%%%%%%%%%%%%%%%%%%%%%%%%%%%%
\begin{figure*}[ht!]
\begin{center}
\includegraphics[width=19cm]{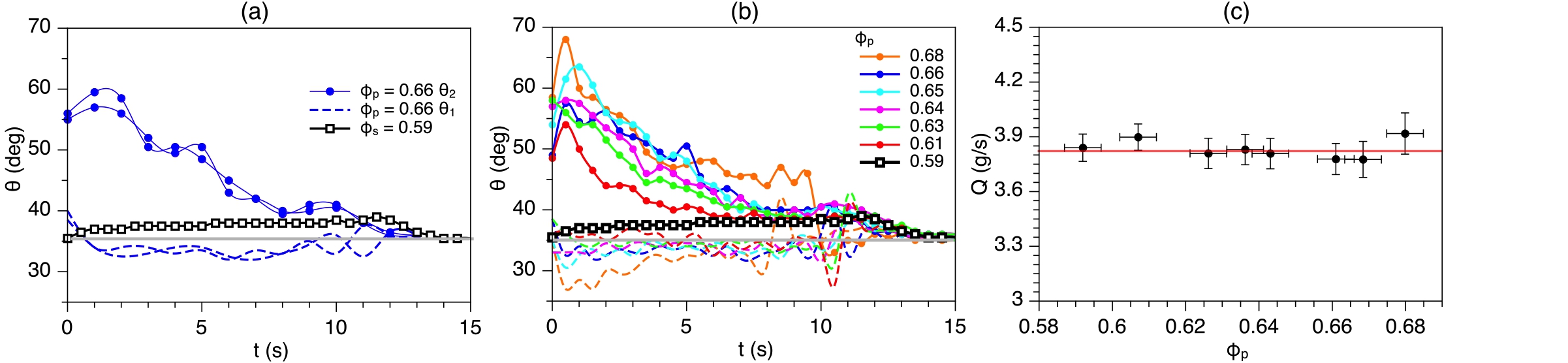}
\caption{Flow rate and evolution of the two surface angles during the discharge: a) $\theta$ vs $t$ for $\varphi_s=0.59$ (-$\square$-) and $\varphi_p=0.66$ (\textcolor{blue}{$--$} $\theta_1$, \textcolor{blue}{$\bullet$} $\theta_2$). b) $\theta$ vs $t$ for different values of $\varphi_p$. Lines with dots and dashed lines of the same colour correspond to $\theta_2$ and $\theta_1$, respectively. The gray line indicates the angle of repose. The angle error due to image analysis was of the order of one pixel, corresponding to $\pm 1^{\circ}$. c) The flow rate $Q=3.82 \pm 0.05$ g/s is constant during the discharge and independent of the initial packing fraction of the bed.} 
\label{fig3}
\end{center}
\end{figure*}
% %%%%%%%%%%%%%%%%%%%%%%%%%%%%%%%%%%%%%%%%%%%%%%%%%%%%%%%%%%%%%%%%%%%%%%%%%%%%%%%

\section{III. Results}

\subsection{A. Conical and double-angle depressions}
Figures \ref{fig1}c-d show a comparison of two-dimensional discharges for standard volume fraction $\varphi_s = 0.59 \pm 0.01$ and for a larger value $\varphi_p = 0.64 \pm 0.01$, respectively. As expected, a surface with a single angle $\theta =37\pm1^{\circ}$ in the standard case is observed, but two angles appear when the packing fraction is increased. By illuminating the system from behind, it is possible to observe the transition from a central zone where the material is flowing and the static lateral zones. The difference in packing produces a sharp contrast in light intensity which allows to define clearly the funnel flow profile. Noteworthy, the transition coincides at the surface level with the point at which the slope changes, indicating an angle dependence on the volume fraction: the inferior angle $\theta_1$ is related to a lower volume fraction in the fluidized zone and the upper angle $\theta_2$ to the larger packing in the static zones. These angles evolve dynamically during the discharge until reaching the same value when the process ends. Figure \ref{fig2} shows the equivalent dynamics in the 3D system: a typical conical depression in standard packing conditions (a), and a depression with two angles for the highly packed case (b). The surface evolution depending on the packing in both 2D and 3D systems can be seen in the Supplementary Movie \cite{SupMovie}.

%%%%%%%%%%%%%%%%%%%%%%%%%%%%%%%%%%%%%%%%%%%%%%%%%%%%%%%%%%%%%%%%%%%%%%%%%%%%%%%
\begin{figure*}[ht!]
\begin{center}
\includegraphics[width=19cm]{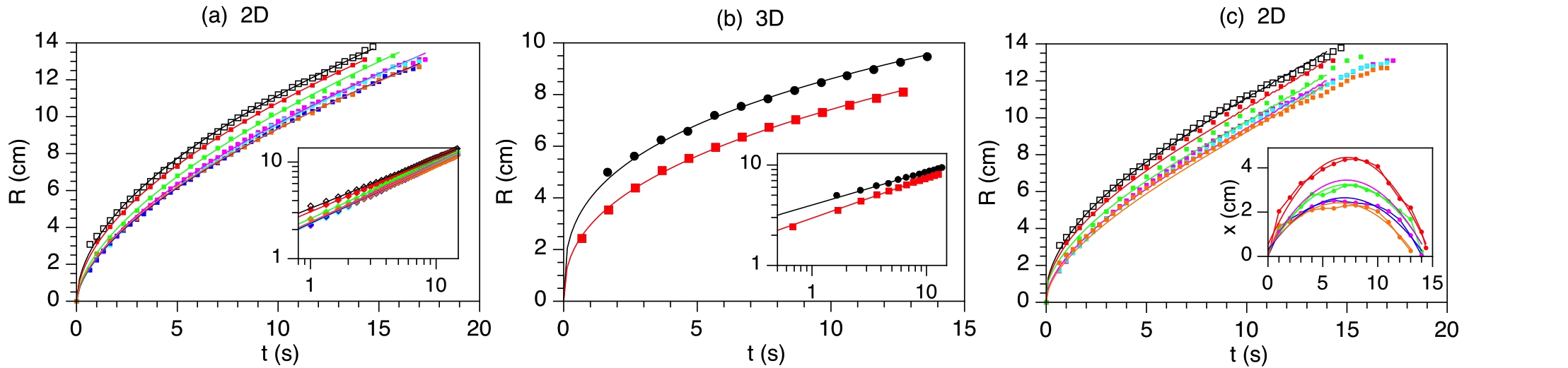}
\caption{Cavity radius as a function of time for a) 2D and b) 3D discharges of standard and densely packed grains; in both cases a power-law dependence is found, see log-log plots in the insets. c) $R$ vs $t$ obtained from the solution of eq. \ref{eqR2} for different values of $\varphi_p$ (dashed lines) compared with the 2D experiments (points); the inset shows the transition coordinate $x(t)$ obtained from the analysis of videos. Colours in (a) and (c) are in correspondence with fig. \ref{fig3}b; in (b): $\varphi_s$ (\textcolor{black}{$\bullet$}) and $\varphi_p$ (\textcolor{red}{$\blacksquare$}).   }
\label{fig4}
\end{center}
\end{figure*}
% %%%%%%%%%%%%%%%%%%%%%%%%%%%%%%%%%%%%%%%%%%%%%%%%%%%%%%%%%%%%%%%%%%%%%%%%%%%%%%%

\subsection{B. Surface evolution and flow rate} The surface angles are easier to measure in the two-dimensional system. Figures \ref{fig3}a-b show the values of $\theta_1$ and $\theta_2$ as a function of time $t$ for different volume fractions $\varphi_p$. In the first plot only the cases $\varphi_s=0.59$ ($\square$) and $\varphi_p=0.66$ (\textcolor{blue}{$\bullet$}) are shown for clarity. Two repetitions exemplify the reproducibility of the process observed in all the events. For the standard case, $\varphi_s$, the slope  fluctuates around the angle of stability $\theta\approx 37.5^{\circ}$ and then falls at the end of the discharge to $\theta_R=35.5^{\circ}$ (gray line), the angle of repose for this granular material. On the other hand, for the close packed bed, $\theta_2$ reaches up to $60^{\circ}$ at the beginning of the discharge while $\theta_1$ falls below the angle of repose, and then both angles converge at the end of the process in $\theta=\theta_R$. In general, the same dynamics is observed for different values of $\varphi_p > \varphi_s$, see fig.\ref{fig3}b: $\theta_2$ increases with $\varphi_p$ (line+symbol) while $\theta_1$ decreases (dashed lines), and always $\theta_1 <\theta_R < \theta_2$. Note that in all cases the discharge finishes practically at the same time $\tau$. Accordingly, fig. \ref{fig3}c shows that the flow rate $Q=\Delta m/\Delta t$ (where $\Delta m$ is  the mass of grains discharged during a time $\Delta t$) measured through the aperture in each case is constant independently of the initial packing conditions of the granular bed. A similar procedure in the 3D case also gives a constant flow $Q=63.7\pm0.9$ g/s for a standard packing ($\varphi_s \approx 0.60\pm 0.01$) and $Q=64.8\pm0.7$ g/s for high-packed conditions ($\varphi_p \approx 0.65\pm 0.01$).

The existence of two angles of discharge produces a cavity with a radial growth rate that depends on $\varphi_p$. Figure \ref{fig4}a shows the radius of the depression $R$ as a function of time $t$ for different values of $\varphi_p$, the inset shows a linear relationship in a log-log plot indicating a power law dependence $R(t)=C(\varphi_p)t^{n(\varphi_p)}$. The standard case, $\varphi_s$, shown in open squares ($\square$) corresponds to the maximum growth rate, and the best fit of the data gives $R(t)_{2D}=(3.32\pm 0.02)t^{0.521\pm0.003}$ (black line). For higher packing values, $R$ grows more slowly over time and the dynamics is well fitted by a power law with smaller coefficients $C(\varphi_p)$ and exponents $n$ that increase with $\varphi_p$ (coloured lines). A similar dependence is obtained from the analysis of a three-dimensional system, see fig. \ref{fig4}b, where the best fit gives $R(t)_{3D}=(3.99\pm0.01)t^{0.333\pm0.005}$ for the standard packing.

%%%%%%%%%%%%%%%%%%%%%%%%%%%%%%%%%%%%%%%%%%%%%%%%%%%%%%%%%%%%%%%%%%%%%%%%%%%%%%%
\begin{figure*}[ht!]
\begin{center}
\includegraphics[width=17cm]{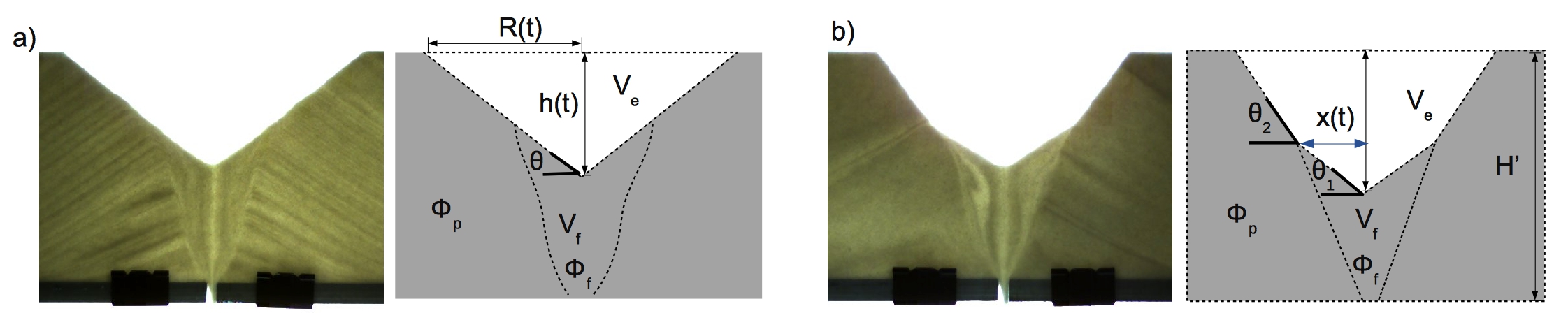}
\caption{Snapshots and diagrams showing the geometry of the cavity for a) standard and b) previously packed systems. $V_e$ indicates the empty volume and $V_f$ the volume occupied by the fluidized material. }
\label{fig5}
\end{center}
\end{figure*}
% %%%%%%%%%%%%%%%%%%%%%%%%%%%%%%%%%%%%%%%%%%%%%%%%%%%%%%%%%%%%%%%%%%%%%%%%%%%%%%%

\subsection{C. Geometrical Model}

Let us now obtain a general expression for $R(t)$ and then consider each particular case (2D, 3D, standard or close-packed bed).  Let $\varphi_p$ and $\varphi_f$ be the volume fractions of the static lateral zones and the fluidized central zone, respectively. From fig. \ref{fig5}, the volumes of the empty depression $V_e$ and the fluidized zone $V_f$ were initially occupied by a mass of grains $\rho \varphi_p  (V_e + V_f)$; this mass must be equal at a given time $t$ to the discharged mass $Q t$ plus the remaining grains in $V_f$ with volume fraction $\varphi_f$; therefore $\rho \varphi_p (V_e + V_f)= Q t + \rho \varphi_f V_f$. Hence we get:
\begin{equation}
\varphi_p V_{e}+(\varphi_p-\varphi_f)V_f=\rho^{-1} Qt
\label{eq1}
\end{equation}

When the material is discharged in standard packing conditions, the difference between $\theta_1$ and $\theta_2$ is negligible and we can assume $\varphi_p \approx  \varphi_f$; thus, the second term in equation (\ref{eq1}) vanishes and $V_e$ in 2D is approximately the volume of the triangular depression indicated in the scheme of fig. \ref{fig5}a; then, $\varphi_p R h w = \rho^{-1}Q t$, where $w$ is the cell width. Since $h=R tan\theta$, we simply obtain:
\begin{equation}
R(t)_{2D} = Ct^{0.5} ,  \hspace{1cm} C=\sqrt{\frac{Q}{\rho \varphi_p w \tan \theta} }
\label{eqR}
\end{equation}
Using the experimental parameters, one gets $C=3.23 \pm 0.09$. This value and the exponent $n=0.5$ obtained from the analysis are in excellent agreement with the best fit of the data shown in fig. \ref{fig4}a for the standard volume fraction. The small difference in exponents can be understood considering that $\varphi_f$ is actually slightly smaller than $\varphi_s$.  

On the other hand, the solution of equation (\ref{eq1}) for larger volume fractions demands an expression for $V_e$ and $V_f$; if $x(t)$ is the x-coordinate of the transition point, see the scheme in fig. \ref{fig5}b, we can approach the volumes $V_e= x^2w(\tan \theta_1 - \tan \theta_2) + R^2w\tan\theta_2$,  and $V_f=w (H'-R\tan\theta_2)x + x^2w(\tan \theta_2 - \tan \theta_1)$, where $R, x, \theta_1$ and $\theta_2$ depend on time; then, equation (\ref{eq1}) can be written as: $\varphi_p R^2 \tan \theta_2 + (\varphi_p-\varphi_f)(H\varphi_f/\varphi_p - R\tan\theta_2)x + \varphi_f x^2(\tan \theta_1 - \tan \theta_2) = \rho^{-1} w^{-1} Qt $, which is a quadratic equation in $R(t)$ with general solution:

\begin{widetext}
\begin{equation}
R(t)_{2D}= -\frac{x}{2}\Big(1 - \dfrac{\varphi_f}{\varphi_p}\Big) 
\pm \Bigg(\Big[\frac{1}{4}\Big(1 - \frac{\varphi_f}{\varphi_p}\Big)^2 + \Big(1 -\frac{\tan \theta_1}{\tan \theta_2}\Big)\frac{\varphi_f}{\varphi_p}\Big]x^2 + \frac{Qt}{\rho \varphi_p w\tan \theta_2} \Bigg)^{0.5}   
\label{eqR2}
\end{equation}
\end{widetext}

Note that eq. \ref{eqR2} reduces to eq. \ref{eqR} for the conical case, where $\varphi_f, \varphi_p \approx \varphi_s $ and $\theta_1 = \theta_2$. To obtain a numerical solution of eq. \ref{eqR2}, we can assume based on fig. \ref{fig3}b that $\theta_1 \approx \theta_R$ and a hyperbolic dependence for $\theta_2 = \theta_0/(1+\beta t)$, where $\theta_0$ is the maximum angle at the beginning of the discharge and $\beta=(\theta_0/\theta_R -1)/\tau$. Moreover, we used a parabolic dependence for $x(t)$ according to measurements of the funnel flow profile shown in the inset of fig. \ref{fig4}c. Under such considerations, one obtains that the positive solutions for $R(t)$ plotted in fig. \ref{fig4}c  for different packing fractions (dashed lines) are in very good agreement with experiments (points). 

Regarding the 3D case, equation (\ref{eq1}) remains valid with a depression described for the standard experiment by a conical shape of volume $V_e= \frac{1}{3}\pi R^2 h$. Considering again $\varphi_p \approx \varphi_f$ and $h = R  \tan \theta$ for this trivial case, one obtains:
\begin{equation}
R(t)_{3D} = Ct^{1/3},  \hspace{1cm} C=(\frac{3Q}{\rho \varphi_p \pi \tan \theta})^{1/3}
\label{eqR3D}
\end{equation}
Using the experimental values one finds $C=3.94\pm0.02$, again, an excellent agreement with experiments (black points) and data fitting (black line) in fig. \ref{fig4}b is obtained. As in the 2D analysis, an analytical expression for larger $\varphi_p$ can be derived based on fig. \ref{fig5}b. In this case, eq. \ref{eq1} can be written as: 
$\varphi_p \tan \theta_2 R^3 + \varphi_f x^3 (\tan \theta_1 - \tan \theta_2) + (\varphi_p -\varphi_f)x^2 (H\varphi_f/\varphi_p  - R \tan \theta_2 )= 3 Q t / \pi \rho$. The third term of the left side is one order of magnitude smaller than the others, therefore, one finds:
\begin{equation}
 R(t)_{3D} \approx \Bigg(\frac{3Qt}{ \rho \varphi_p \pi \tan \theta_2} + \frac{\varphi_f}{\varphi_p}(1-\frac{\tan \theta_1}{\tan \theta_2})x^3\Bigg)^{1/3} 
\label{eqR3D-2}
\end{equation}

As it is to be expected, eq. \ref{eqR3D-2} is reduced to eq. \ref{eqR3D} when $\theta_1 = \theta_2$. To solve eq. \ref{eqR3D-2}, it is necessary to determine the complex dependence of $x$ and $\theta_2$ on time, but these functions cannot be measured directly as it was possible in the 2D case; nonetheless, our analysis in 2D and 3D show that in all cases $R(t)$ can be well approximated by a power law of the form $R(t)=C(\varphi_p)t^{n(\varphi_p)}$, where $C(\varphi_p)$ includes the complex dependence on $t$ found in eqs. 3  and 5.  \\

 %%%%%%%%%%%%%%%%%%%%%%%%%%%%%%%%%%%%%%%%%%%%%%%%%%%%%%%%%%%%%%%%%%%%%%%%%%%%%%%
\begin{figure*}[ht!]
\begin{center}
\includegraphics[width=17.5cm]{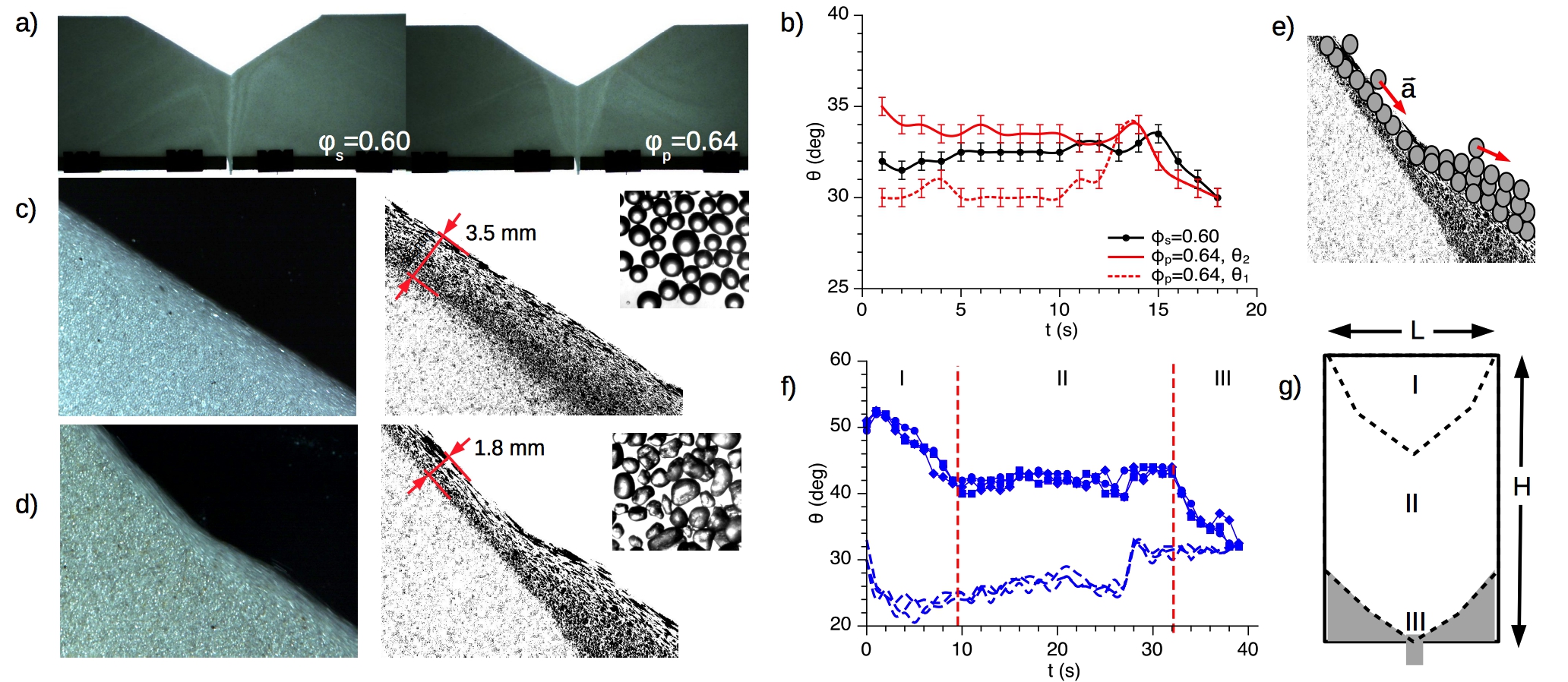}
\caption{a) Discharge of the 2D cell with glass beads for $\varphi_s =0.60$ and $\varphi_p = 0.64$. b) Evolution of surface angles as a function of $t$ during de discharge of glass beads. c-d)  Close-up view of the flowing layer for glass beads and sand grains, respectively. The thickness of the moving layer is indicated in each case. e) Scheme of particles flowing from the lateral zone to the fluidized zone, where the grains accumulate. f) Evolution of $\theta_1$ and $\theta_2$ as a function of $t$ during the discharge of a 2D tall silo, where $H>L$. g) Stages of the tall silo discharge according to (f). Each level ends when: (I) the cavity reaches the side walls, (II) the cavity reaches the outlet, and (III) the flow stops at the angle of repose.  }
\label{fig6}
\end{center}
\end{figure*}
% %%%%%%%%%%%%%%%%%%%%%%%%%%%%%%%%%%%%%%%%%%%%%%%%%%%%%%%%%%%%%%%%%%%%%%%%%%%%%%%

\subsection{D. Rough grains vs smooth grains}

Although a high volume fraction seems to be necessary to obtain two angles of discharge, smooth spherical glass particles packed at the same conditions apparently develop only one surface angle independently of $\varphi_p$, see fig. \ref{fig6}a. A thorough analysis of the surface profile at the transition zone reveals a slope difference $\theta_2 - \theta_1 \approx 4^{\circ}$, see fig. \ref{fig6}b. This small change contrasts with the difference of up to $ 30^{\circ}$ during the discharge of sand grains for the same packing fraction. Both, sand grains and glass beads have approximately the same density ($\rho\approx 2.65\pm 0.02$ g/cc) and average size ($d \approx 200$ $\mu$m), the main difference between them is the particle shape and roughness, as it can be noticed from the snapshots in figs. \ref{fig6}c-d. A close up view of the surface depression in both cases reveals that the flowing layer of glass beads is almost twice thicker than the layer of sand, while they are approximately of the same thickness in standard conditions. Another visible difference is that the transition from the moving layer to the static zone is sharp for sand and disperse for glass beads. On the other hand, regardless the packing conditions, the flow rate for glass beads was $Q=3.86\pm0.04$ g/s, approximately the same value obtained with sand grains.

\section{VI. CONCLUSIONS AND DISCUSSION}

According to our observations, the surface profile can be useful to determine the packing conditions of the granular bed. For a standard volume fraction, the difference between the maximum angle of stability in the static zone and in the fluidized zone is small  and it cannot be distinguished at a glance, but a larger contrast between these angles is a clear indication of a high-packed material and it can be useful to prevent silo collapse.  Moreover, the greatest change is observed with rough amorphous grains, and this scenario is the most common and closest to large scale processes if we consider that edible grains in agriculture, cement, gravel and other materials in industry are indeed rough non-spherical particles. The results presented here are in agreement with previous studies about the stability of heaps made of natural granular materials\cite{Arias2011}, where it was found that less spherical particles have greater stability angles. Similar results were obtained using rotating drum systems, showing an important increase of $\theta_m$ in two-dimensional heaps of trapezoids and diamond-shaped grains compared to more rounded geometries\cite{Olson2002,Olson2005}. It was also found in the last reference  that for most of the geometries the heap stability increases with packing fraction, which is also consistent with our findings.

Two external angles of stability have also been recently reported in numerical investigations of very large three-dimensional heaps of particles produced by ballistic deposition\cite{Topic2012}. In that research, it was found that distinct density zones produced during the deposition have a remarkable implication in the angle of repose, and the angle change was associated with an increase on the number of contacts among beads. This coincides with our explanation of the notable increase of angle for non-spherical particles: amorphous grains can be rearranged under vibration to maximize the contact area, in contrast to glass beads whose contact is reduced to a surface point; therefore, a grain of sand experiences more friction with the surrounding particles and needs on average a greater angle of inclination to start moving.

Another aspect to notice is that the relation  $\theta_1 <\theta_R< \theta_2$ is always satisfied. Since $\theta_1$ is the angle corresponding to the fluidized zone with volume fraction $\varphi_f$, its value should be independent on $\varphi_p$; nevertheless, $\theta_1$ slightly decreases when the initial packing of the bed is augmented. To explain this fact, let us remember that the fluidized zone is fed by the thin layers of grains flowing over the static zones. Considering an average friction coefficient $\mu$, these grains have an acceleration $a \sim g(\sin \theta_2 - \mu \cos \theta_2)$, where $g$ is the acceleration of gravity. Thus, the velocity at which they arrive to the fluidized zone is greater for higher values of $\theta_2$ and they can move further to be accumulated near the central zone, see sketch in Fig. \ref{fig6}e. The accumulation of particles generates a slope with $\theta_1< \theta_R$ and it allows the avalanche to stop. Furthermore, fig. \ref{fig3} shows that $\theta_2$ decreases over time. This is also related to the accumulation of grains in the central zone and only happens if the flow from the lateral zones is greater than the flow  through the outlet. As the discharge time $t$ increases, the distance from the cavity rim to the fluidized zone augments and the upper angle must diminish from geometrical requirements. Thus, $\theta_2$ would not change if such distance were constant. To test this point, we performed experiments in a tall cell with $H > (L/2)\tan\theta_R$, see sketch and data in figs. \ref{fig6}f,g. As expected, $\theta_2$ remains constant during a considerable part of the discharge (stage II) because the length of the flowing layer over the stagnant zone is practically the same in such interval. 

Finally, it is important to argue about the independence of $Q$ on the packing fraction.  Even though the flow rate in vertical silos filled under different conditions has been investigated\cite{Ahn2008}, small flow-rate variations are related to changes in pressure and not in the packing fraction itself, and little effect on the flow is expected in vertical silos packed by the influence of gravity\cite{Aguirre2014}. Our results shown in fig. \ref{fig3}c clearly indicate a constant $Q$ for different values of $\varphi_p$. This fact can be understood if we observe from snapshots in fig. \ref{fig1}c-d that the material from the stagnant zone must pass through the fluidized zone before reaching the aperture; therefore, the initial volume fraction of the bed is irrelevant. Indeed, a dependence of $Q$ on the value of $\varphi$ only has been found important in vertical\cite{Huang2006} and horizontal silos\cite{Aguirre2014} in a regime of dilute flows of particles, where a larger concentration at the outlet produces a larger amount of grains leaving the system. Then, it could be of interest for future research to explore the flow rate transition from dilute to dense flows of granular materials.  

\section*{Acknowledgments}
This research was supported by the following projects: CONACyT Mexico No. 242085 of the Sectoral Research Fund for Education, PRODEP-SEP No. DSA/103.5/14/10819 and PROFOCIE-SEP 2015-2016.

Corresponding author: fpacheco@ifuap.buap.mx

\bibliographystyle{apsrev}

\bibliography{Biblio-BB}

\end{document}